\documentclass[final,5p,times,twocolumn]{elsarticle}

\usepackage{lipsum}
\usepackage{lineno}
\usepackage{amssymb}
\usepackage{amsmath}
\usepackage{here}
\usepackage{braket}
\usepackage{bm}
\usepackage[colorlinks=true, bookmarks=true,bookmarksnumbered=true,bookmarkstype=toc]{hyperref}
\usepackage{braket}
\usepackage[pdftex]{xcolor}
\usepackage{physics}

\usepackage{multirow}
\usepackage{colortbl}
\usepackage{cancel}

\journal{}

\begin{document}

\begin{frontmatter}

\title{Measurement of the charge-to-mass ratio of particles trapped by the Paul trap for education}
\author[tohoku]{R. J. Saito}
\ead{ryuta.saito.q3@dc.tohoku.ac.jp}
\author[utokyo]{T. A. Tanaka}
\author[cns]{Y. Sakemi}
\author[cns]{M. Yagyu}
\author[waseda]{K. S. Tanaka}

\address[tohoku]{Department of Physics, Tohoku University, 6-3 Aramaki-aza Aoba, Aoba-ku, Miyagi 980-8578, Japan}
\address[utokyo]{Graduate School of Arts and Sciences, The University of Tokyo, 3-8-1 Komaba,Meguro-ku, Tokyo 153-8902, Japan}
\address[cns]{Center for Nuclear Study, Graduate School of Science, The University of Tokyo, 2-1 Hirosawa, Wako, Saitama 351-0198, Japan}
\address[waseda]{Waseda Research Institute for Science and Engineering, Waseda University, 3-4-1 Okubo, Shinjuku-ku, Tokyo, 169-8555, Japan}

\begin{abstract}
Paul traps are devices that confine particles using an alternating electric field and have been used in undergraduate experimental classes at universities. Owing to the requirement of a high voltage ($> 10^3$ V), Paul traps are not used in middle and high schools. Therefore, we developed an all-in-one-type Paul trap , including a high-voltage transformer. The Paul trap can be equipped with two different types of electrode attachments, ring-type and linear-type, and the trap image can be observed using a built-in web camera. For example, the charge-to-mass ratio of particles was measured with different types of attachments, and reasonable values were obtained. These types of trap devices are currently used at several educational facilities in Japan.
\end{abstract}

 
\end{frontmatter}

\section{Introduction}
A Paul trap uses an alternating electric field to confine charged particles in space. Simple Paul trap devices are generally used in undergraduate experiments conducted at several universities.
\par
The Tokyo Institute of Technology fabricated a device using machined copper electrodes and used it in undergraduate experiments\cite{kobayashi1996}. This device could generate a deep electric potential close to the ideal rotational hyperbolic shape and trap a few particles near the center of the potential.
\par
Simple devices that can be fabricated include a trapping apparatus at Caltech \cite{caltech}, which employs metal ring electrodes. A device, including its electrodes, can be created using a three-dimensional (3D) printer, which is made available online by S'CoolLAB \cite{doi:10.5334/joh.12}. S'CoolLAB's trap can be assembled using commercially available parts or via modeling using a 3D printer; it can observed Coulomb crystals owing to its expansive trappable area and ability to confine multiple particles.
\par
However, in all such cases, preparing a high-voltage power supply of several thousand volts is necessary, and thus, junior high and high school students are unable to operate such devices.
\par
We developed an all-in-one-type Paul trap that can be operate using the standard household power supply and a web camera for observation and recording. By placing the high-voltage parts inside the case, it can be safely used for demonstrations and student experiments in junior high and high schools.
\par
\begin{figure}[htbp]
    \begin{center}
    \includegraphics[width=\hsize,keepaspectratio]{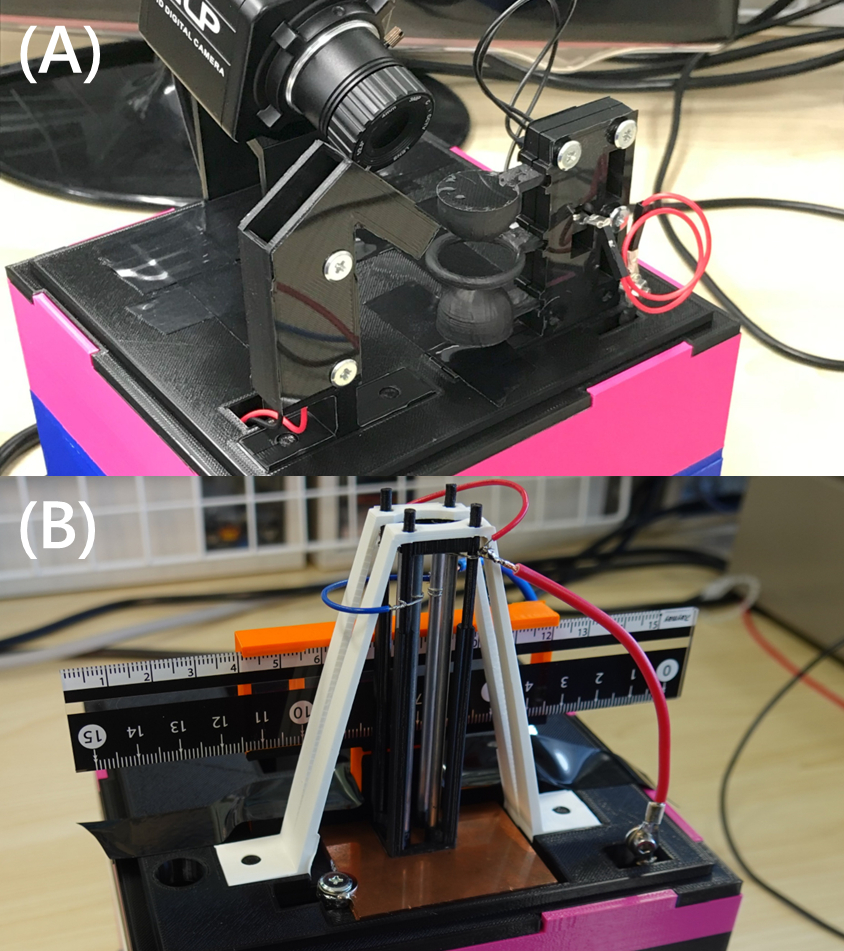}
    \end{center}
    \caption{Photograph of two Paul trap modules. (A) ring-type configuration, (B) linear-type configuration.}
    \label{fig:paultrap-photo}
\end{figure}

This Paul trap device has two replaceable electrodes: ring-type and linear-type electrodes (Fig.~\ref{fig:paultrap-photo}). In this study, the charge-to-mass ratio of trapped particles was measured in three different ways to test the feasibility of the Paul trap as experimental equipment in high schools.
\begin{enumerate}
   \item measurement of applied AC high voltage conditions that can trap particles using the ring-type Paul trap.
   \item balance between the gravity and electrostatic force generated by the DC electric field using the linear-type Paul trap.
   \item measurement of the amplitude of forced vibrations generated by applying an external oscillation voltage using the linear-type Paul trap.
\end{enumerate}

\section{Principle of the Paul trap}
According to the Earnshaw's theorem, a charged particle cannot be confined statically by electrostatic forces alone, and an alternating field is required to trap it. 
A positive particle is trapped by an electrostatic field with a harmonic potential expressed as  
\begin{align*}
    \phi = \alpha x^2 + \beta y^2 + \gamma z^2,
\end{align*}
The potential must satisfy the Poisson equation as
\begin{gather*}
    \nabla \phi = 0 \\
    \therefore \alpha + \beta + \gamma = 0.
\end{gather*}
One among $\alpha$, $\beta$, and $\gamma$ must be negative; that is, the restoring force will not operate in a certain direction and will not be able to trap. Therefore, an alternating electric field must be used to trap the particles.
\par
Two types of Paul traps were used in this experiment: (1)a Paul trap that generates a rotating hyperbolic potential with $\alpha = \beta$, $\gamma = -2\alpha$ (ring-type setup) and (2) a Paul trap that confines only in the $x$- and $y$-directions with $\alpha = -\beta$, $\gamma = 0$ (linear-type setup.).
\par
\begin{figure}[htbp]
    \begin{center}
    \includegraphics[width=\hsize,keepaspectratio]{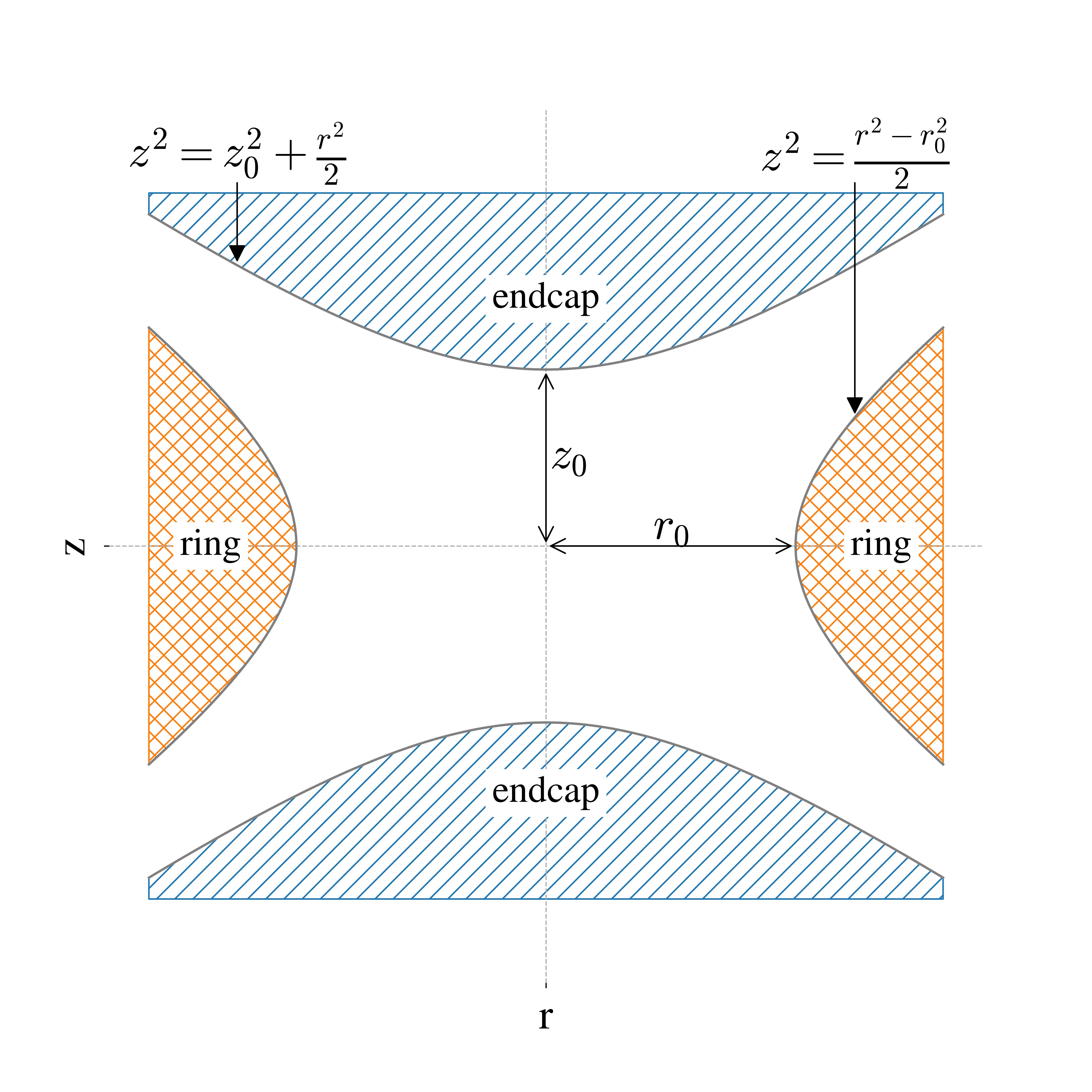}
    \end{center}
    \caption{Schematic of a ring-type Paul trap electrode. There is a relationship $r_0^2 = 2z_0^2$ between $r_0$ and $z_0$.}
    \label{fig:ring-electrode-schema}
\end{figure}
\begin{figure}[htbp]
    \begin{center}
    \includegraphics[width=\hsize,keepaspectratio]{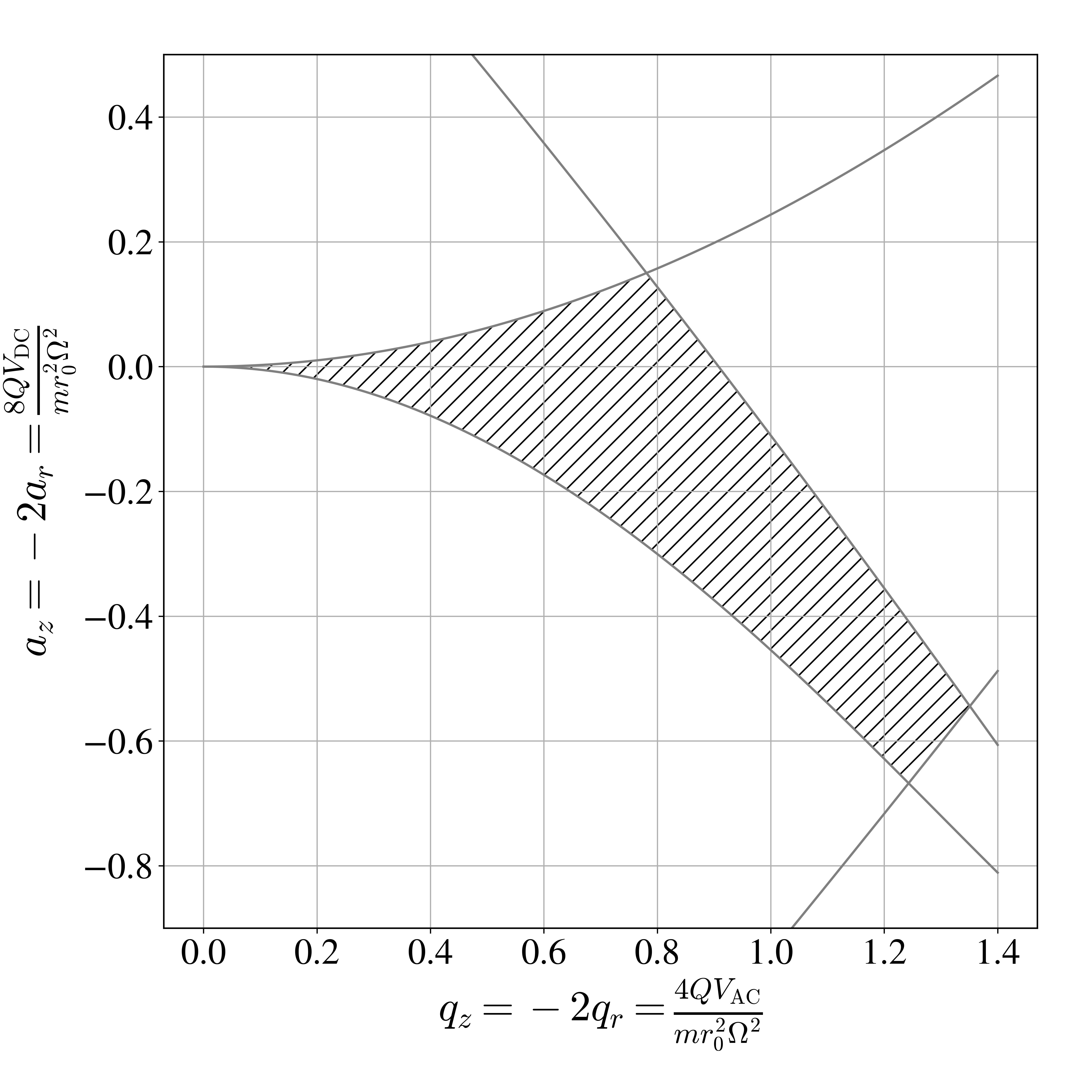}
    \end{center}
    \caption{Stability diagram for Mathieu equation. The shaded area is the stability region.}
    \label{fig:diagram}
\end{figure}

The motion of the charged particles in the Paul trap obeys the Mathieu equation. Next, we consider the equations of motion for trapped particles of mass $m$ and charge $Q$; the electrode shown in Fig.~\ref{fig:ring-electrode-schema} is a rotational hyperbolic potential
\begin{gather}
    \phi = \frac{\phi_0}{2r_0^2}(x^2 + y^2 - 2z^2) = \frac{\phi_0}{2r_0^2}(r^2 - 2z^2) \label{eq:potential}\\
    \phi_0 = V_{\rm DC} + V_{\rm AC} \cos (\Omega t). \notag
\end{gather}
Here, $V_{\rm DC}, V_{\rm AC}$ are the DC and AC voltages applied to the electrodes, respectively; $\Omega$ is the angular frequency of the AC voltage. The equation of motion for the trapped particles is
\begin{alignat*}{2}
    r &\colon &\hspace{2pt} m\dv[2]{r}{t} &= -\frac{Q}{r_0^2} \left[V_{DC} + V_{AC} \cos (\Omega t) \right] r\\
	z &\colon &\hspace{2pt} m\dv[2]{z}{t} &= \frac{2Q}{r_0^2} \left[ V_{DC} + V_{AC} \cos (\Omega t) \right] z.
\end{alignat*}
using the dimensionless quantity $\tau = \Omega t/2$. 
\begin{gather}
    \dv[2]{u}{\tau} + \left[a_u - 2q_u \cos(2\tau) \right]u = 0\quad (u = r, z) \label{eq:mathieu}\\
    a_z = -\frac{Q}{m}\frac{8V_{\rm DC}}{r_0^2\Omega^2 } = -2a_r \notag\\
	q_z = \frac{Q}{m}\frac{4V_{\rm AC}}{r_0^2\Omega^2} = -2q_r, \notag
\end{gather}
where $a, q$ are dimensionless quantities that depend on the charge-to-mass ratio of the charged particles. Equation~\ref{eq:mathieu} is the Mathieu Equation and is characterized by the values of the dimensionless quantities $a, q$ that determine the stability of the trap (Fig.~\ref{fig:diagram}). 

\par

\begin{figure}[htbp]
    \begin{center}
    \begin{minipage}{0.48\linewidth}
        \centering
        \includegraphics[width=\hsize, keepaspectratio]{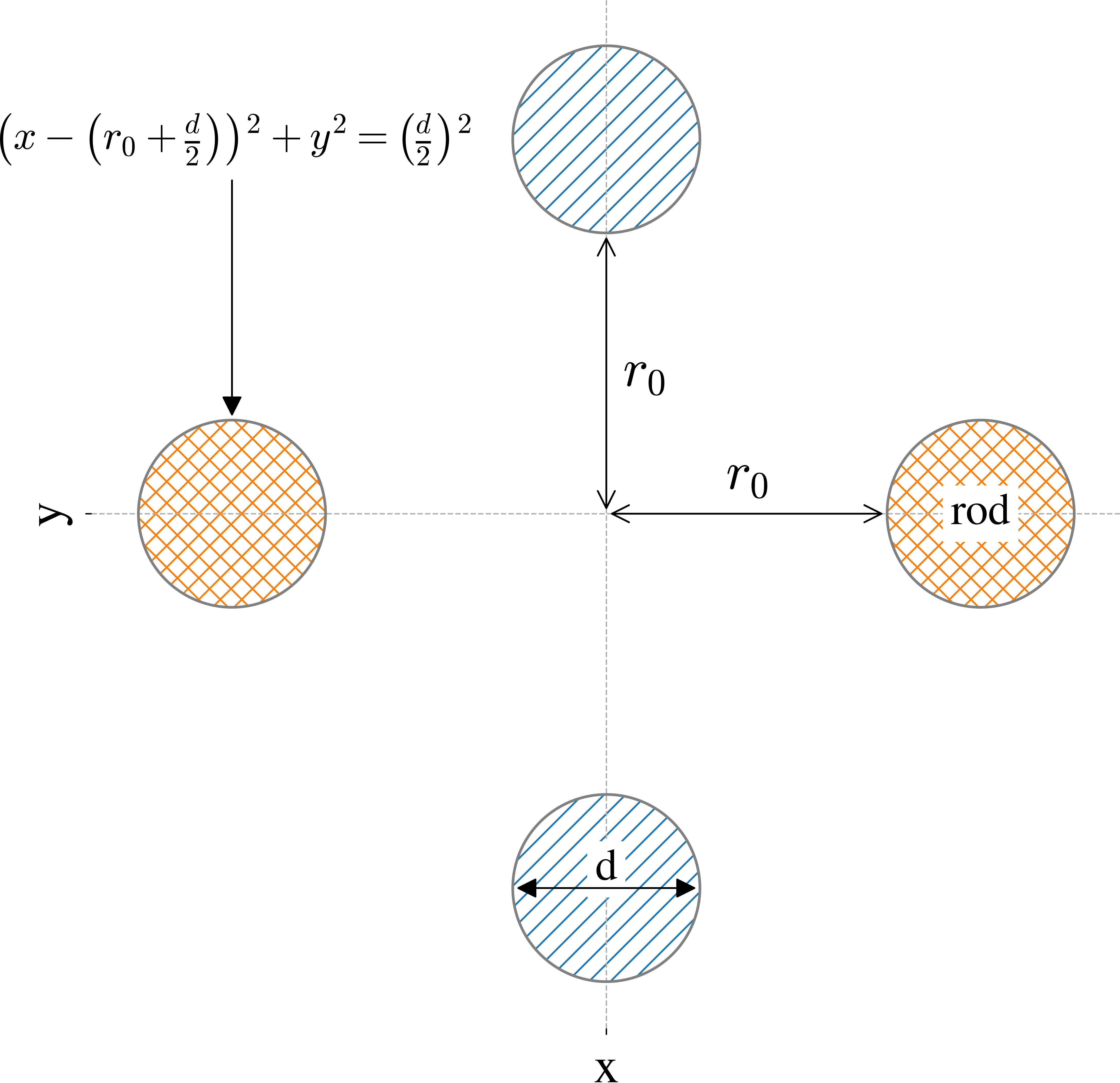}
    \end{minipage}
    \begin{minipage}{0.48\linewidth}
        \centering
        \includegraphics[width=\hsize, keepaspectratio]{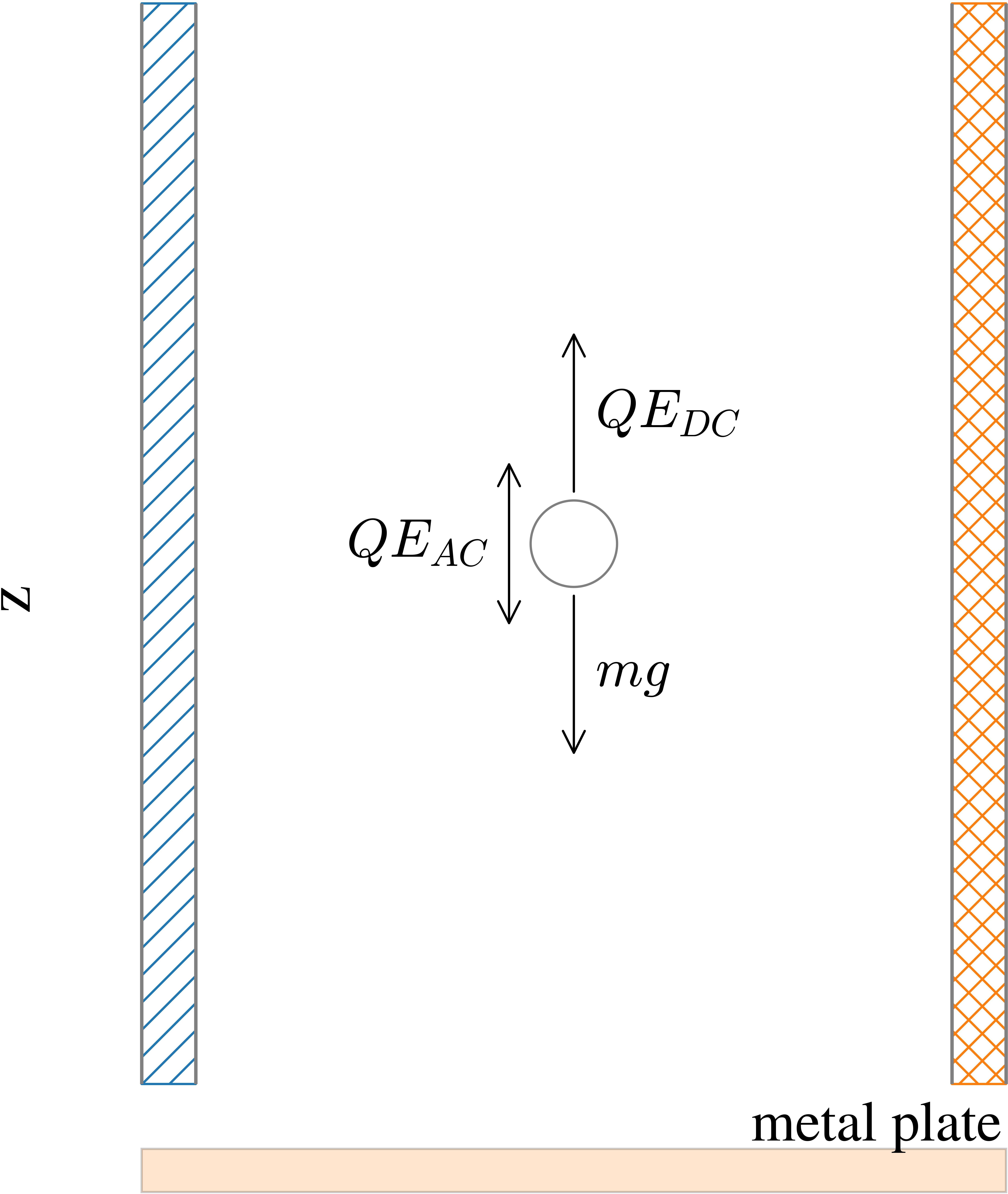}
    \end{minipage}
    \end{center}
    \caption{Schematic of a linear-type Paul trap electrode. $E_{\rm DC}$ and $E_{\rm AC}$ denote the electric fields created by the bottom plate.}
    \label{fig:linear-electrode-schema}
\end{figure}

In the case of the linear-type Paul trap (Fig.~\ref{fig:linear-electrode-schema}), equation~\ref{eq:potential} is modified by substituting $r$ with $x$ and $z$ with $y/\sqrt{2}$, such as
\begin{align*}
    \phi = \frac{\phi_0}{2r_0^2}(x^2 - y^2).
\end{align*}
The corresponding results are essentially equivalent to those obtained for the ring-type Paul trap.
\par

\par

 The viscous and inertial forces of air should be considered when particles are trapped in the air. The size of the Lycopodium spores \cite{caltech} used as trapped particles in this experiment was 
\begin{align*}
    D = 2R &= 26.0 \pm 2.5\ {\rm \mu m}\\
    \rho &= 510 \pm 40\ {\rm kg/m^3}.
\end{align*}
Therefore, the Reynolds number $R_e$ of Lycopodium spores is
 \begin{align*}
    R_e = \frac{\rho_{\rm air} vD}{\eta} \approx 1.7 \times 10^{-4} \ll 1
\end{align*}
where the density of air is $\rho_{\rm air} = 1.21\ {\rm kg/m^3}$, dynamic viscosity is $\eta = 1.82 \times 10^{-5}\ {\rm Pa \cdot s}$ and velocity of the particle is $v = 10^{-5}$ to $10^{-3}\ {\rm m/s}$. Thus, the viscous force is dominant over the inertial force in the case of these spores. Therefore, Eq.~\ref{eq:mathieu}, with the coefficient of the viscous resistance as $k$, is
\begin{gather}\label{eq:damping-mathieu}
    \dv[2]{u}{\tau} + b\dv{u}{\tau} + \left[a_u - 2q_u \cos(2\tau) \right]u = 0\quad (u = r, z)\\
    b = \frac{2k}{m\Omega}. \notag
\end{gather}
Furthermore, assuming that the particles are spherical, the proportionality coefficient $k$ can be rewritten as $6\pi \eta R$ from Stokes' law, and the mass of the particles as $m = 4/3\pi R^3 \rho$, the dimensionless quantity $b$ is
\begin{align*}
    b = \frac{9\eta}{R^2 \rho \Omega}.
\end{align*}

\section{Designing the Paul trap}
\begin{figure}[htbp]
    \begin{center}
    \includegraphics[width=\hsize,keepaspectratio]{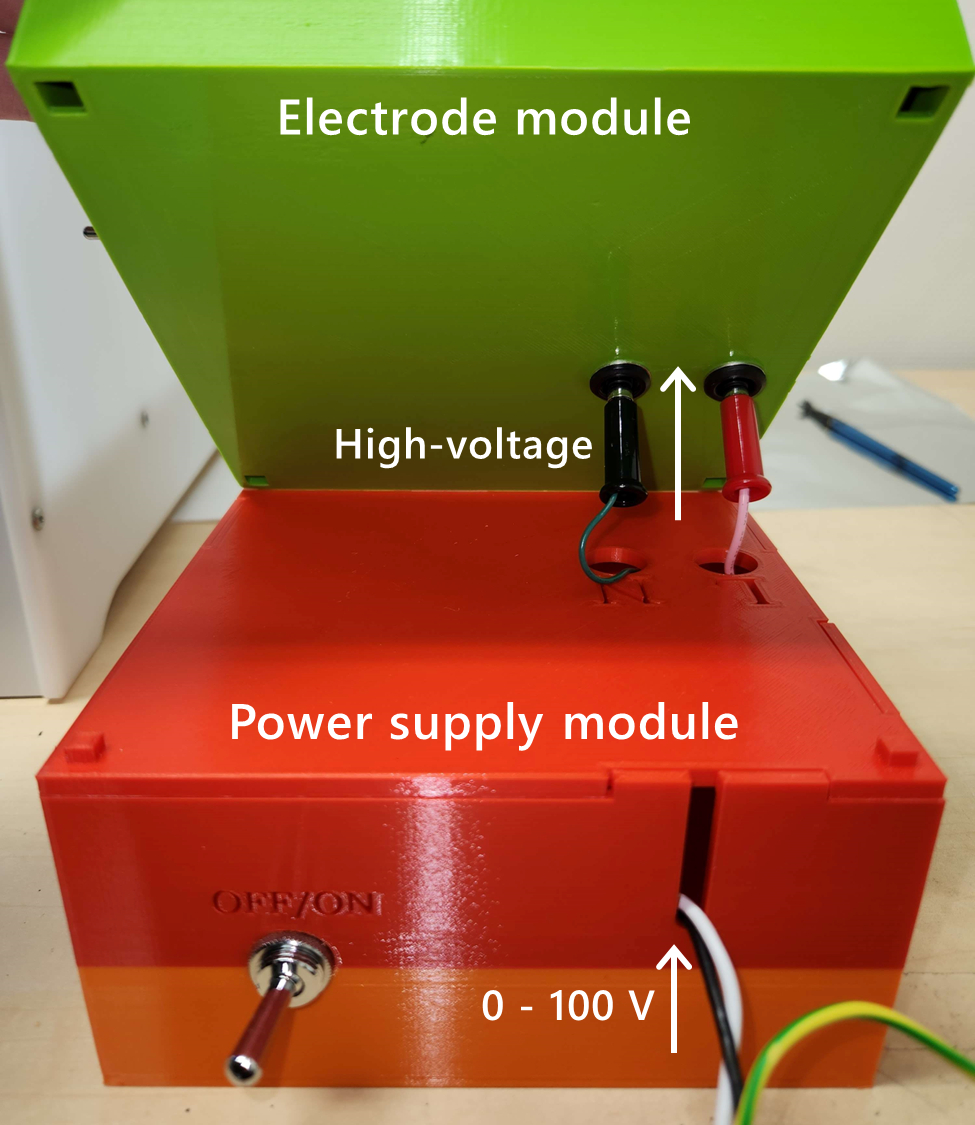}
    \end{center}
    \caption{The upper box represents the electrode module, and the lower box contains the power supply module with a high-voltage transformer. The power supply module is shared, allowing for different electrode configurations by replacing the upper box for varying the experimental setups.}
    \label{fig:device}
\end{figure}

The device was divided into two components: a power supply module and an electrode module (Fig.~\ref{fig:device}). The power supply module contained a 60-fold high-voltage transformer (UFO-6K-001-P100, UNION ELECTRIC) and amplified the voltage from that of a household power supply in Japan (100 V / 50 or 60 Hz) to 6000 V. The electrode modules (ring-type and linear-type) were switchable for each measurement. Each part was fabricated using a 3D printer (Ultimaker S3) and polylactic acid (PLA) filament. The 3D-printable models, along with a simple assembly manual, are publicly available on GitHub \cite{model_github}. The trapped particles were observed using a mounted USB webcam (USB130W01MT-MF40-J).
\par
\begin{figure}[htbp]
    \begin{center}
    \includegraphics[width=\hsize,keepaspectratio]{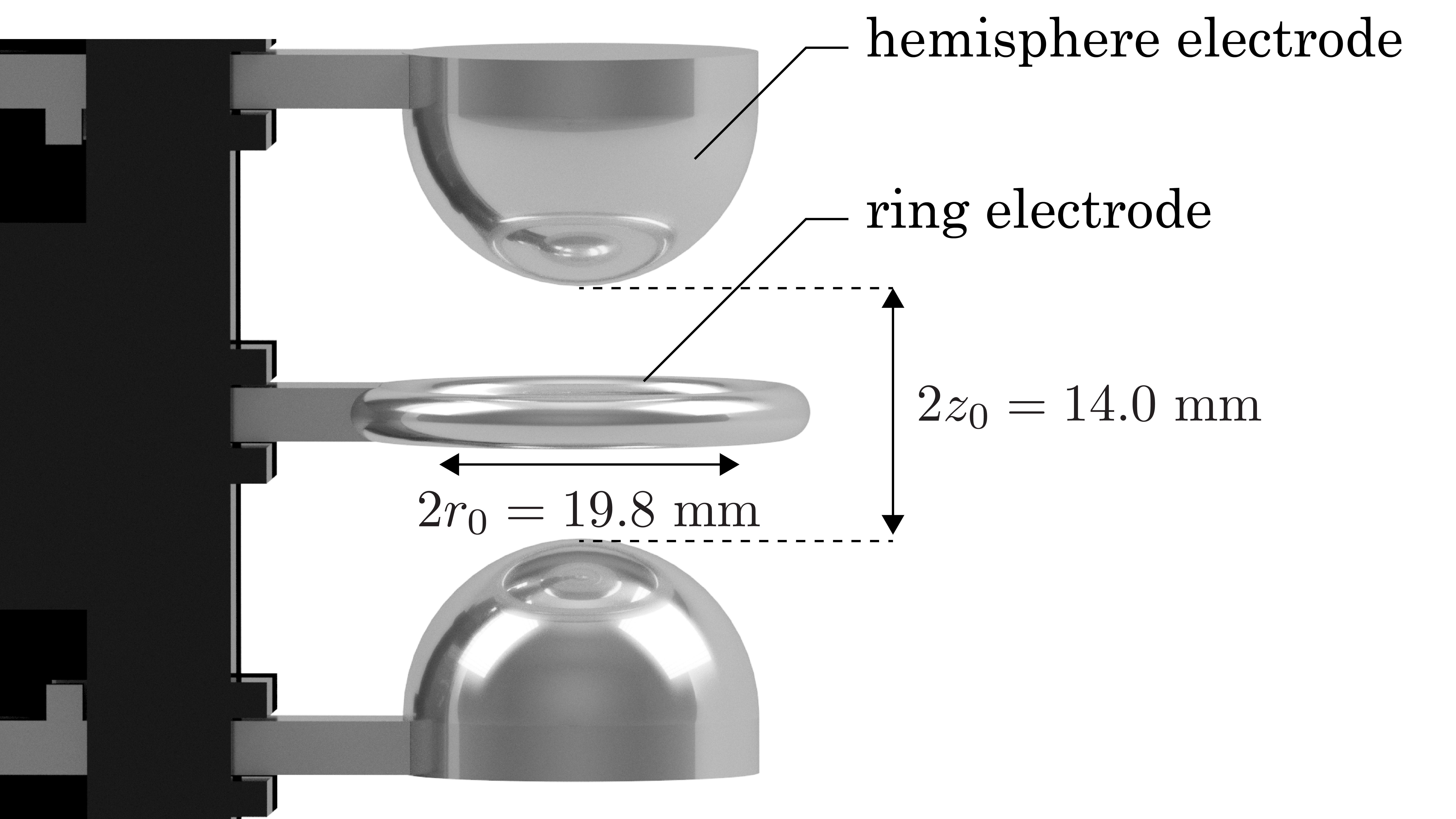}
    \end{center}
    \caption{3D model of attachable electrodes of the ring-type Paul trap.}
    \label{fig:ring-3D}
\end{figure}

The ring-type setup comprised two hemispheres and a ring electrode (Fig.~\ref{fig:ring-3D}). The electrodes were 3D printed, and a conductive paint (SKU-0216, BARE CONDUCTIVE) was applied to enable voltage application. This type applied only an AC high voltage ($ V_{\rm DC} = 0 $); hence, $a = 0$ (experiment 1).
\par
\begin{figure}[htbp]
    \begin{center}
    \includegraphics[width=\hsize,keepaspectratio]{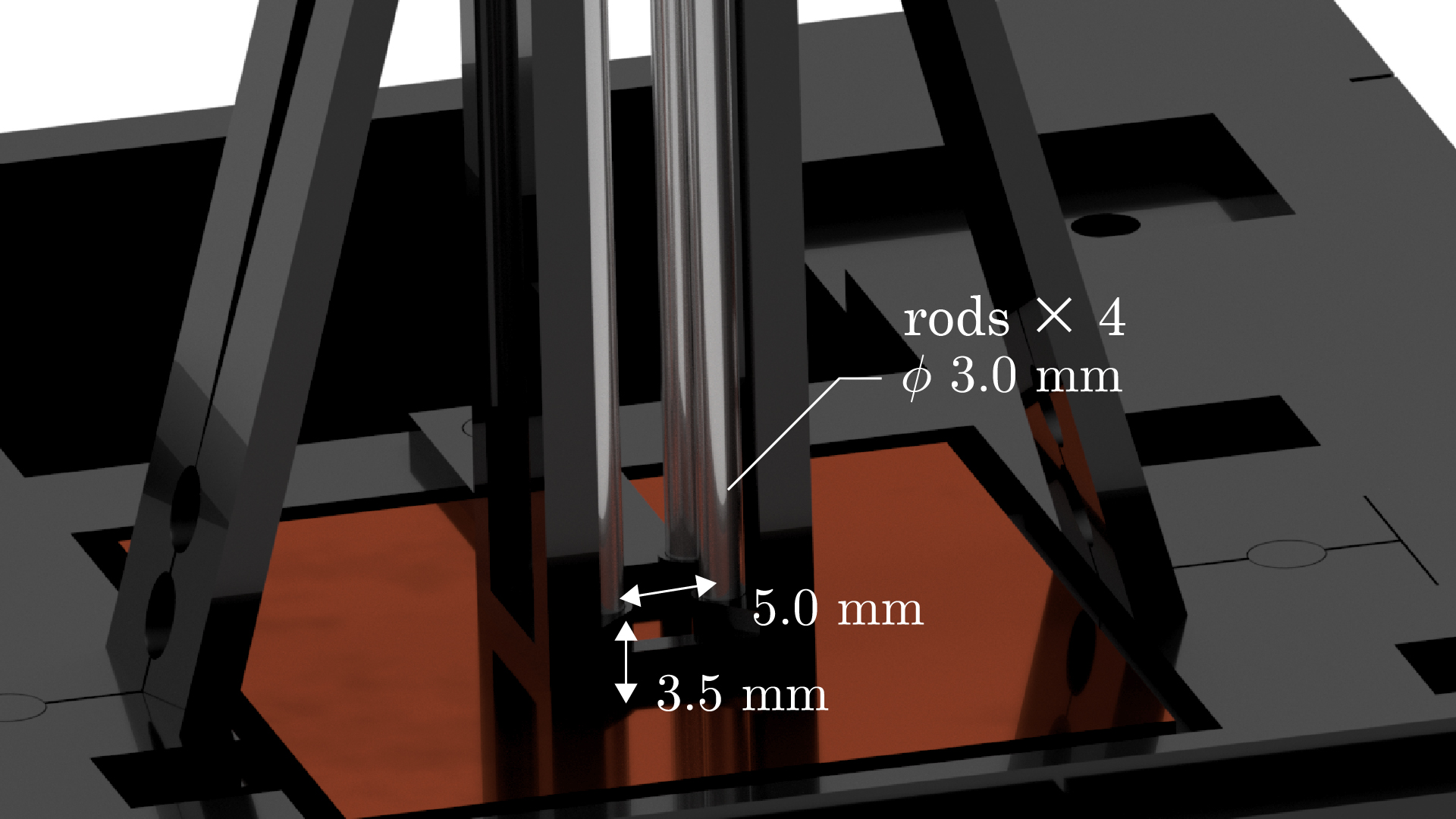}
    \end{center}
    \caption{3D model of attachable electrodes of linear-type Paul trap.}
    \label{fig:linear-3D}
\end{figure}

The linear-type setup comprised four metal rods for trapping a particle via AC high voltage application; it also contained a bottom plate for DC voltage (experiment 2) or DC and AC voltage (experiment 3) application (Fig.~\ref{fig:linear-3D}).
\par
\begin{figure}[htbp]
    \begin{center}
    \includegraphics[width=\hsize,keepaspectratio]{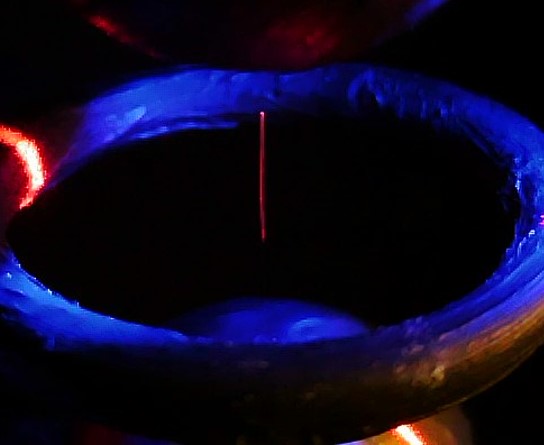}
    \end{center}
    \caption{Observed trajectory of particles. Due to the refresh rate of the webcam (30 Hz), the trajectory is seen as a line.}
    \label{fig:trap-single-paerticle}
\end{figure}

Lycopodium spores, which are easily available spherical plant particles, were used as trap particles (Fig.~\ref{fig:trap-single-paerticle}).

\section{Measurement of charge-to-mass ratio}
\subsection{Applied AC high voltage conditions that can trap particles using the ring-type Paul trap}
According to Eq.~\ref{eq:mathieu} and \ref{eq:damping-mathieu}, the trap condition is determined by $q$ and air resistance coefficient $b$.
By obtaining the maximum values of $q$ and $V_{\rm AC}$ under stable trapping conditions ($q_{\rm{max}}$ and $V_{\rm AC}^{\rm max}$), the charge-to-mass ratio of the particles $Q/m$ can be calculated using the following equation:
\begin{align*}
    \frac{Q}{m} = \frac{r_0^2 \Omega^2}{4} \frac{1}{V_{\rm AC}^{\rm max}} q_{\rm max}.
\end{align*}
\par
 $q_{\rm{max}}$ is calculated using the Runge-Kutta method as shown in Table~\ref{tab:particle_param} and the following equation, which considers the gravitational force in Eq.~\ref{eq:damping-mathieu}
\begin{align*}
    &\dv[2]{r}{\tau} + b\dv{r}{\tau} + \left[a_r - 2q_r \cos (2\tau)\right]r = 0\\
    &\dv[2]{z}{\tau} + b\dv{z}{\tau} + \left[a_z - 2q_z \cos (2\tau)\right]z + \frac{4g}{\Omega^2} = 0.
\end{align*}
The calculated $q_{\rm{max}}$ is: 
\begin{align*}
    q_{\rm{max}} =  17 \pm 4
\end{align*}
\begin{table}[htbp]
    \caption{Parameters to calculate the stability condition of the ring-type setup.}
    \begin{center}
    \begin{tabular}{c|c}
    \hline
    Parameter &Value\\
    \hline
    \hline
    $r_0$ & $9.9 \pm 0.5\ {\rm mm}$\\
    $\Omega$ & $2\pi \times 50\ {\rm Hz}$\\
    $\eta$ & $ 1.81920(6) \times 10^{-5}\ {\rm Pa\cdot s}$\\
    $R$ & $13.0 \pm 1.3\ {\rm \mu m}$\\
    $\rho$ & $510 \pm 40\ {\rm kg/m^3}$\\
    \hline
    \end{tabular}
    \label{tab:particle_param}
    \end{center}
\end{table}
\par
A single particle was trapped with the ring-type Paul trap, and we gradually raised the applied voltage $V_{\rm{AC}}$ to escape. The obtained maximum $V_{\rm{AC}}$ under the trap conditions is
\begin{align*}
   V_{\rm{AC}}^{\rm max} = 2.5 \pm 0.8\ \rm{kV}.
\end{align*}
From $q_{\rm{max}}$ and $V_{\rm AC}^{\rm max}$, $Q/m$ can be calculated as:
\begin{align*}
    \frac{Q}{m} =  (1.7 \pm 0.7) \times 10^{-2}\ \rm{C/kg}.
\end{align*}

\subsection{Gravity--Electrostatic forces equilibrium under DC electric field using the linear-type Paul trap}
The gravitational and electrostatic forces from the bottom plate are balanced in the $z$ direction with the linear-type Paul trap as
\begin{align*}
    mg = QE_{\rm DC}
\end{align*}
Thus, the charge-to-mass ratio is calculated as follows:
\begin{align} \label{eq:balance}
    \frac{Q}{m} = \frac{g}{E_{\rm DC}}.
\end{align}
\par
\begin{figure}[htbp]
    \begin{center}
    \includegraphics[width=\hsize,keepaspectratio]{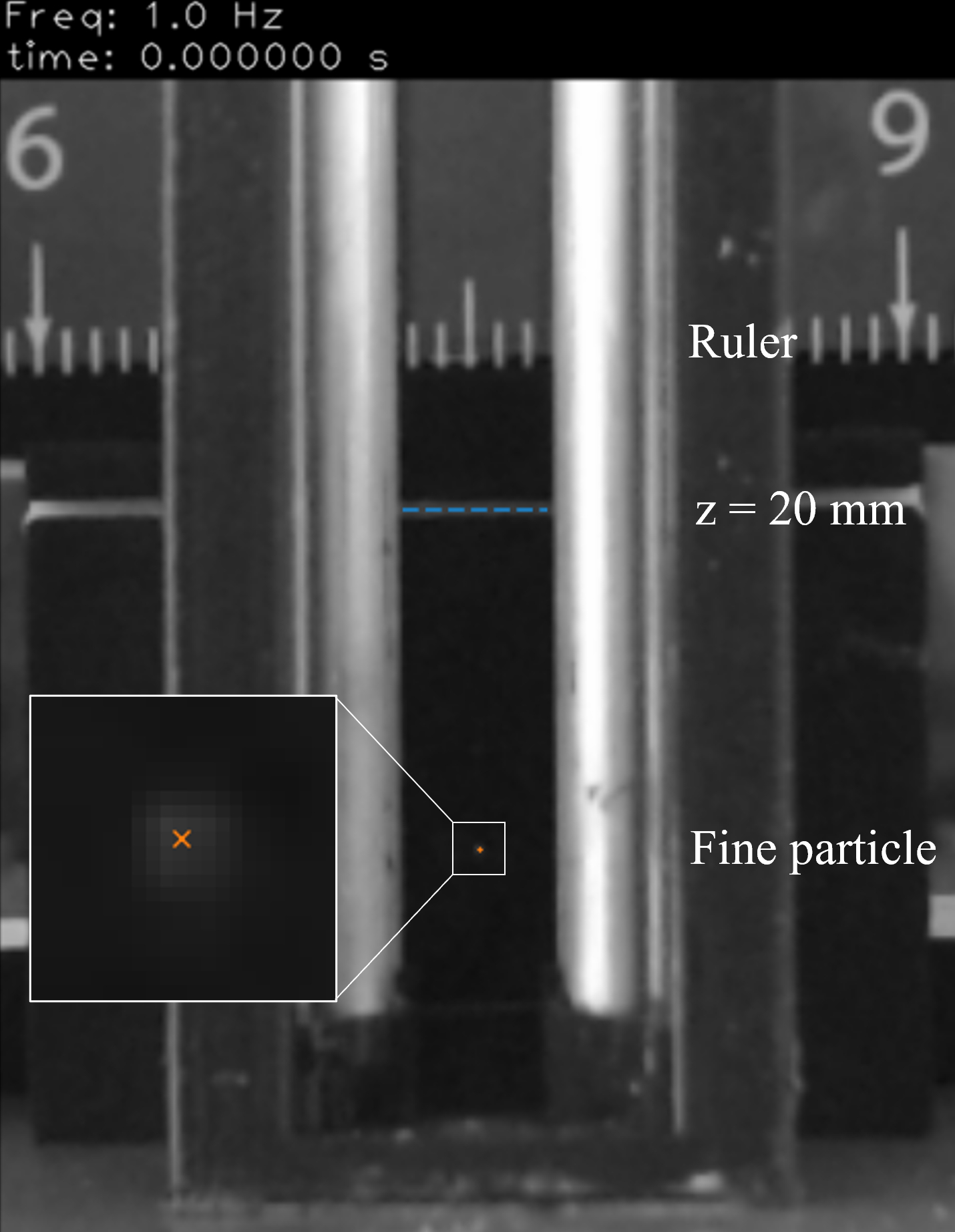}
    \end{center}
    \caption{Observed image of a particle trapped by the linear-type Paul trap. Positions of the particle are measured with reference to the line ($z = 20$\ mm) and ruler in this view. }
    \label{fig:vertical-linear-analysis}
\end{figure}

A single particle was trapped in the linear-type Paul trap under the condition of $V_{\rm{AC}}^{\rm trap}$ = 2100 V, $V_{\rm{DC}} = -80$ to $-150$\ V where $V_{\rm{DC}}$ is static applied voltage to the bottom plate. Subsequently, particle displacement was measured according to the increase in $V_{\rm{DC}}$ (Fig.~\ref{fig:vertical-linear-analysis}). 
\par
\begin{figure}[htbp]
    \begin{center}
    \includegraphics[width=\hsize,keepaspectratio]{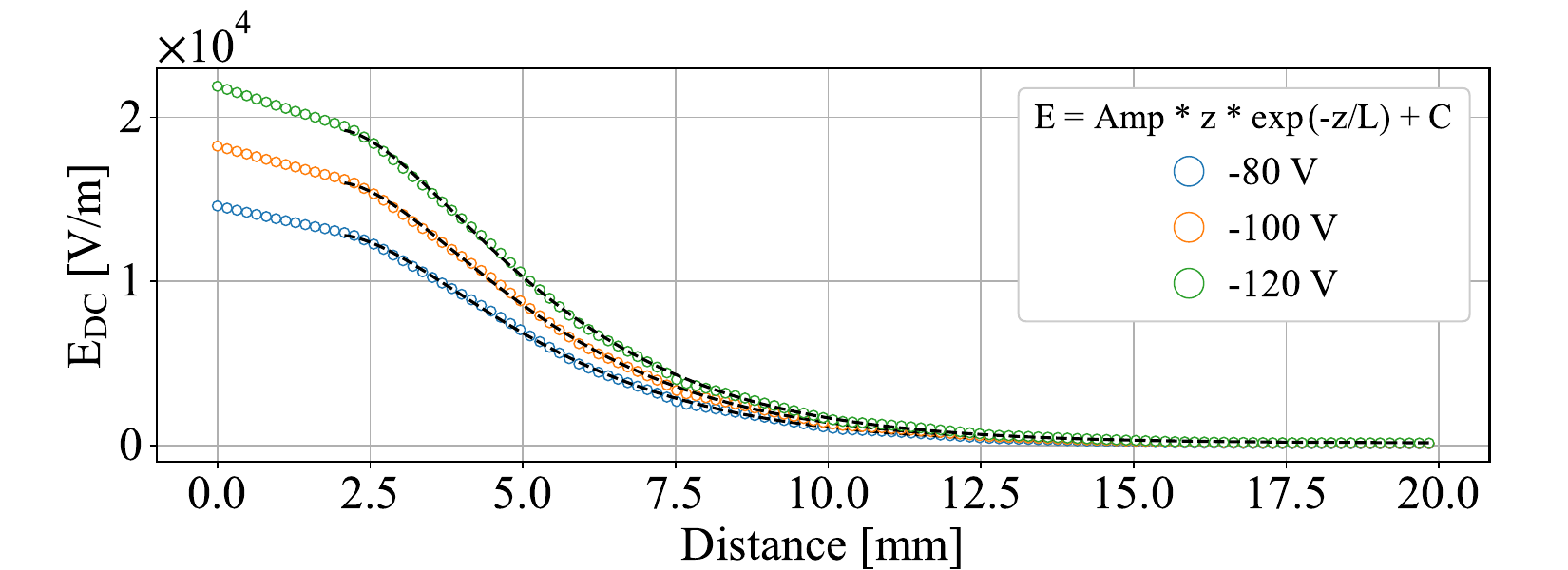}
    \end{center}
    \caption{Electric field strength on the $z$ axis with applied static voltage $E_{\rm{DC}} = -80, -100$, and $-120$ V calculated using the CST Studio. The distribution on the $z$ axis was fitted by the exponential function $E_{\rm{DC}}(z) = E_0 z \exp \qty(-z/L) + C$ to obtain the electric field strength at each particle's position.}
    \label{fig:Esimu-DC}
\end{figure}

The distribution of the static electric field was calculated using the Statics / Low-frequency Solver available in the CST Studio Suite, and we fitted the distribution on the $z$ axis using an exponential function, given by $E_{\rm{DC}}(z) = E_0 z \exp \qty(-z/L) + C$, to determine the electric field strength at each particle's position (Fig.~\ref{fig:Esimu-DC}).

\begin{figure}[htbp]
    \begin{center}
    \includegraphics[width=\hsize,keepaspectratio]{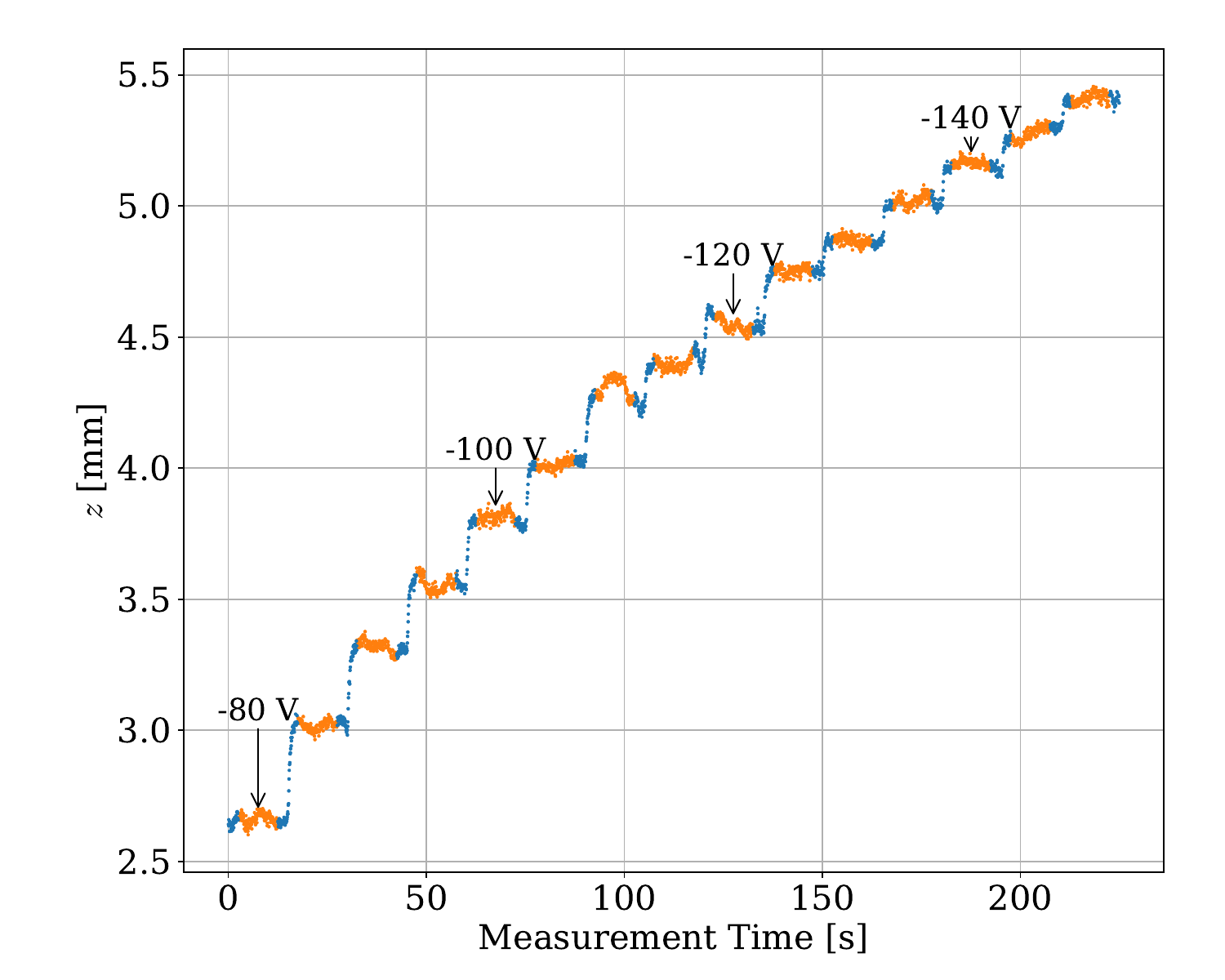}
    \end{center}
    \caption{Position change of the particle by the applied static voltage $V_{\rm{DC}}$.}
    \label{fig:Vdc-pos}
\end{figure}

\begin{figure}[htbp]
    \begin{center}
    \includegraphics[width=\hsize,keepaspectratio]{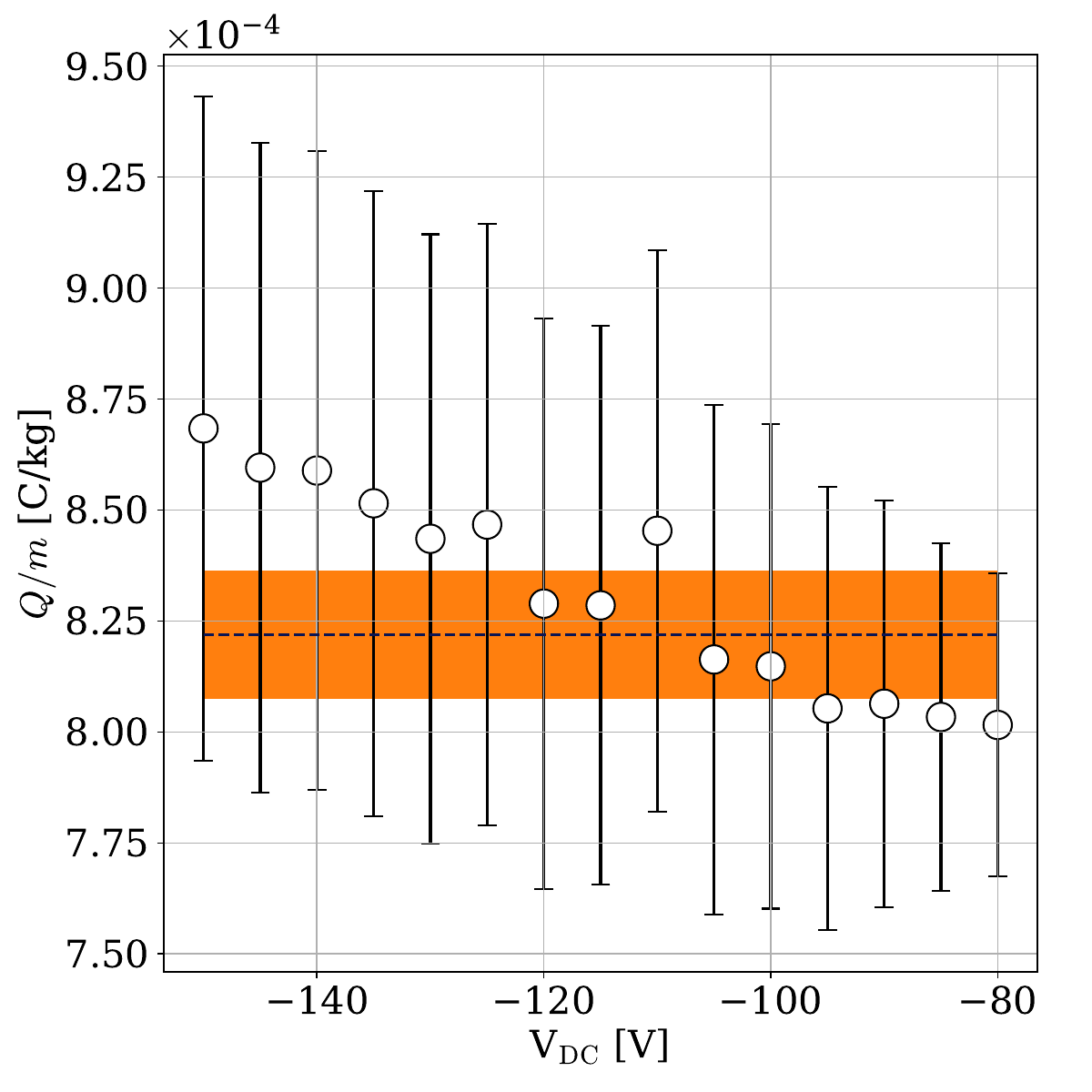}
    \end{center}
    \caption{Calculated charge-to-mass ratio ($Q/m$) by each applied static voltage $V_{\rm{DC}}$.}
    \label{fig:Vdc-qm}
\end{figure}

Figure~\ref{fig:Vdc-pos} shows the relationship between the particle displacement and $V_{\rm{DC}}$.
The charge-to-mass ratio ($Q/m$) calculated using Eq.~\ref{eq:balance} for each $V_{\rm{DC}}$ (Fig.~\ref{fig:Vdc-qm}) and the weighted average value is 
\begin{align*}
    \frac{Q}{m} =  (8.2 \pm 0.1) \times 10^{-4} \ {\rm C/kg}.
\end{align*}

\subsection{Amplitude of forced vibrations generated by applying an external oscillation voltage using the linear-type Paul trap}
\begin{figure}[htbp]
    \begin{center}
    \includegraphics[width=\hsize,keepaspectratio]{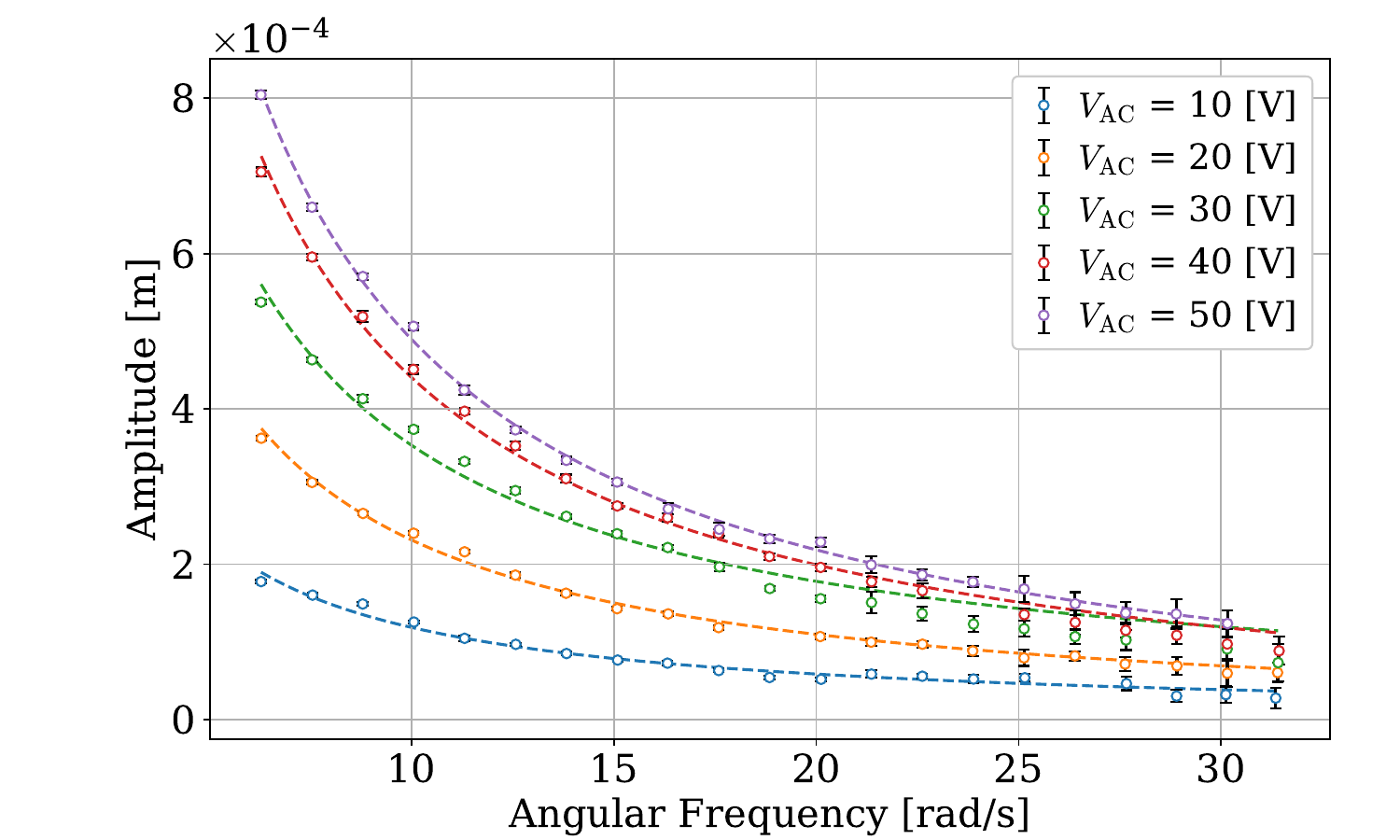}
    \end{center}
    \caption{Amplitudes of forced vibrations with several applied oscillation voltages $V_{\rm{AC}}$ and frequencies. Amplitudes are inversely proportional to the frequency of applied oscillation fields according to Eq.~\ref{eq:Vac-amp}.}
    \label{fig:Vac-amp}
\end{figure}

\begin{figure}[htbp]
    \begin{center}
    \includegraphics[width=\hsize,keepaspectratio]{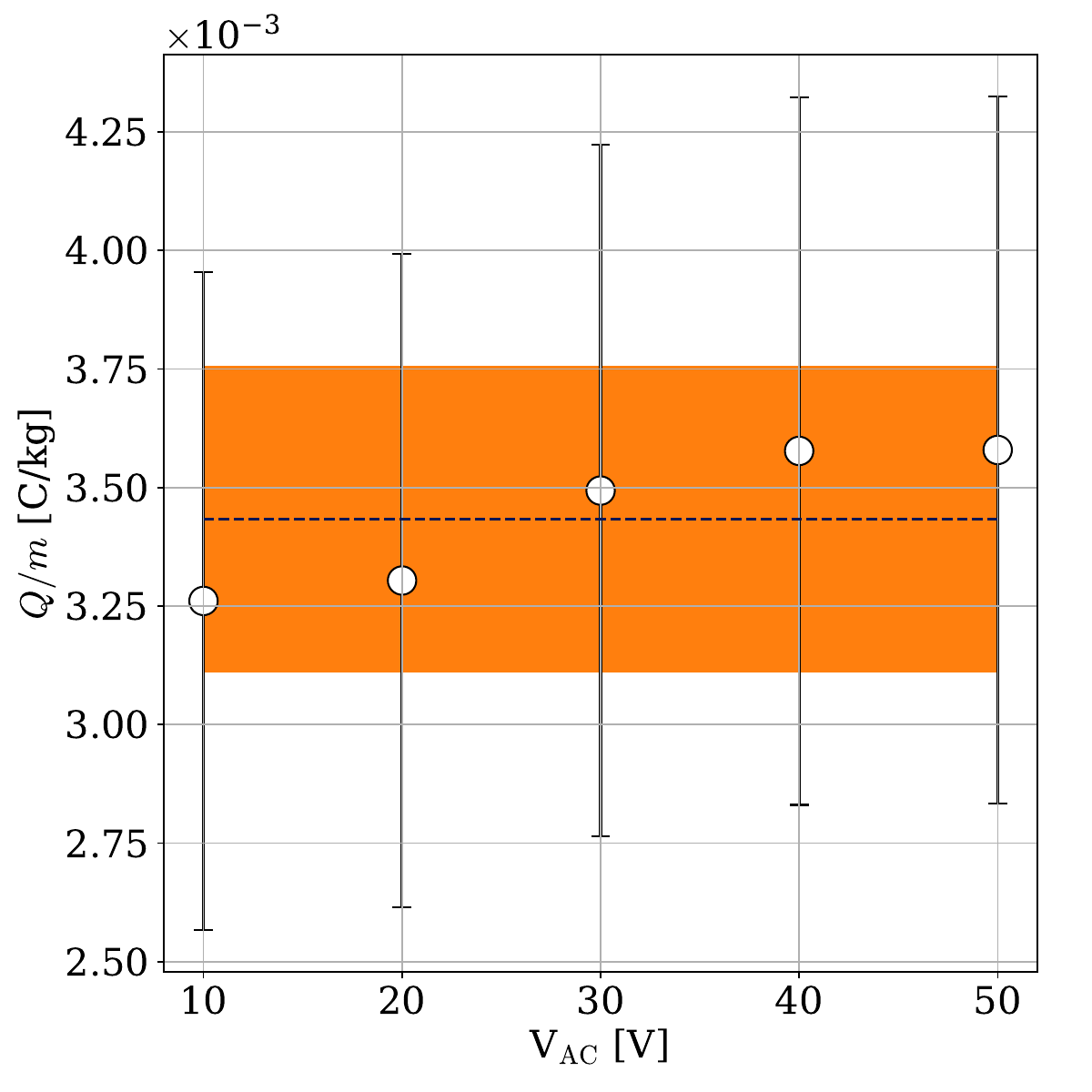}
    \end{center}
    \caption{Calculated charge-to-mass ratio ($Q/m$) by each applied oscillation voltage $V_{\rm{AC}}$.}
    \label{fig:Vac-qm}
\end{figure}

A single particle was trapped in the linear-type Paul trap under the condition of $V_{\rm AC}^{\rm trap}$ = 2100 V and $V_{\rm{DC}} = -80$ V. Furthermore, an additional oscillating electric field ($E_{\rm AC} \sin(\omega t)$, where $E_{\rm AC}$ is a function of $V_{\rm AC}$), applied to the bottom plate at a low (sub-Hz) frequency induced forced vibrations as
\begin{align*}
    m\dv[2]{z}{t} &= -k\dv{z}{t} + QE_{\rm AC} \sin(\omega t) + \cancel{QE_{\rm DC} - mg},\\
    z &= A\sin(\omega t + \phi)
\end{align*}
where the amplitude $A$ is expressed as follows:
\begin{align}
    A &= \frac{QE_{\rm AC}}{k\omega} \left[ 1 + \qty(\frac{m\omega}{k})^2 \right]^{-\frac{1}{2}}\notag \\
    &\approx \frac{Q}{m}\frac{mE_{\rm AC}}{k}\frac{1}{\omega} \quad \qty(\because \frac{m\omega}{k} \ll 1)\notag  \\
    &= \frac{Q}{m}\frac{2R^{2}\rho E_{\rm AC}}{9\eta}\frac{1}{\omega}.  \label{eq:Vac-amp}
\end{align}
Thus, the charge-to-mass ratio ($Q/m$) can be obtained by measuring $A$ for each $\omega$. Figure~\ref{fig:Vac-amp} shows the measured $A$ values for $V_{\rm{AC}}=10, 20, 30, 40$, and $50$ V. The data measured for each $V_{\rm AC}$ is fitted by Eq.~\ref{eq:Vac-amp}, resulting in the determination of the charge-to-mass ratio (Q/m) for each $V_{\rm AC}$, as depicted in Figure~\ref{fig:Vac-qm}. Subsequently, the weighted average of these individual values is calculated as
\begin{align*}
    \frac{Q}{m} = (3.4 \pm 0.3) \times 10^{-3}\ \rm{C/kg}.
\end{align*}

\section{Conclusions}
In this study, we developed a tabletop Paul trap that could be replaced with two attachable traps, ring- and linear-type, and measured the charge-to-mass ratio of the particles trapped using three different techniques. This tabletop Paul trap could be operated only on the household power supply, and by connecting the built-in camera to a PC via USB, measurements and image analysis could safely be performed in a typical school environment. Therefore, these three charge-to-mass ratio measurements could be safely performed in an experimental school class.
\par
The results varied in the range of $Q/m = 10^{-3}$ to $10^{-2}$ C/kg, consistent with the typical value for charged Lycopodium spores \cite{caltech}. The uncertainties in the value originated from variations in the spore size and structure as well as the charging method, which involved rubbing a Teflon rod with a cloth and then attaching the spores to it.
\par
For the ring-type Paul trap charge-to-mass ratio measurements (experiment 1), it was necessary to know the particle size and density, and to determine maximum value of the trap stability condition $q_{\rm{max}}$ from numerical calculations. To measure the charge-to-mass ratio in the linear-type Paul trap (experiment 2 and 3), determining the distribution of the electric field on the axis with respect to the applied voltage was crucial. Additionally, for experiment 3, the particle size and density should be determined in advance. Because the electric field distribution could be calculated in advance at the time of production, experiment 2 was found to be the simplest method for measurement the charge-to-mass ratio of particles.
\par
To date, trap devices have been widely used in Japan, such as in a class at Waseda University Honjo Senior High School, booth exhibition at the Sendai Science Museum, and open-campus exhibition at the University of Tokyo, albeit mainly for high school students. In particular, at Waseda University Honjo Senior High School, traps have been used in demonstration classes conducted by high school teachers and in year-long exploratory activities conducted by students.

\section{Acknowledgement}
This research was partially supported by the Mitsubishi Memorial Foundation for Educational Excellence, the Cyclotron Radioisotope Center at Tohoku University, and the Leave a Nest grant incu $\cdot$ be award.

\bibliographystyle{junsrt}
\bibliography{paultrap-en-rev}

\end{document}